 \documentclass[aps,prd,floatfix,onecolumn]{revtex4}

\usepackage[english]{babel}

\usepackage{graphicx}
\usepackage{dcolumn}
\usepackage{bm} 
\usepackage{epstopdf}

\newcommand{\sigi}{\sig_i}

\newcommand{\minfty}{{-\infty}}

\newcommand{\meen}{{ -1}}

\newcommand{\Is}{I_s}

\newcommand{\sig}{\sigma}

\newcommand{\hi}{{(i)}}

\renewcommand{\xi}{\delta N}

\newcommand{\hs}{{(s)}}

\newcommand{\myskip}[1]{}

\renewcommand{\d}{{\rm d}}

\newcommand{\SA}{{\rm SA}}

\newcommand{\BEQ}{\begin{eqnarray}}
\newcommand{\EEQ}{\end{eqnarray}}
\newcommand{\BEA}{\begin{eqnarray}}
\newcommand{\EEA}{\end{eqnarray}}
\newcommand{\nn}{\nonumber}
\newcommand{\p}{\partial}

\newcommand{\sqt}{\sqrt{2}}
\newcommand{\sqf}{\sqrt{5}}

\begin{document}




\begin{center}
{\Large \textbf{Models for quantum measurement of particles with higher spin}}
\end{center}

\begin{center}
Theodorus Maria Nieuwenhuizen
\end{center}

\begin{center}
{Institute for Theoretical Physics,   University of Amsterdam 
 \\ PO Box 94485, 1090 GL  Amsterdam, The Netherlands} \\   
* t.m.nieuwenhuizen@uva.nl
\end{center}

\section*{Abstract}
The Curie-Weiss model for quantum measurement describes the dynamical measurement of a spin-$\frac{1}{2}$ 
by an apparatus consisting of an Ising magnet of many spins $\frac{1}{2}$ coupled to a thermal phonon bath.
To measure the $z$-component $s=-l,-l+1,\cdots,l$ of a spin $l$, a class of models is designed along the same lines, which involve $2l$ order parameters.
 As required for unbiased measurement, the Hamiltonian of the magnet, its entropy and the interaction Hamiltonian
 possess an invariance under the permutation $s\to s+1$ mod $2l+1$.  
The theory is worked out for the spin-1 case, where the thermodynamics is analyzed in detail, 
and for spins $\frac{3}{2},2,\frac{5}{2}$ the thermodynamics and the invariance is presented.  
 

 


 \tableofcontents

  \renewcommand{\thesection}{\arabic{section}}
\section{Introduction}
  \setcounter{equation}{0} \setcounter{figure}{0}
\renewcommand{\thesection}{\arabic{section}.}

The interpretation of quantum mechanics has long been shrouded in mystery.
The best working formulation involves the Copenhagen postulates, while various other attempts are summarized in ref. \cite{wheeler2014quantum}.
While a plethora of (semi-) philosophical papers have been written on the subject,
the one and only touchstone between the quantum formalism and the reality in a laboratory lies in quantum measurement,
hence this connection has been the focus of our research in the last decades.

 Indeed, while indispensable for introductory classes in quantum mechanics,
``Copenhagen'' skips over the reality of a real apparatus performing a measurement in a laboratory, 
and thus bypasses the physics to which it pretends to provide interpretation.
It is best seen as a short cut to the reality of measurement, useful for introductory courses on quantum mechanics, but
lacking rigour at a fundamental level.

What is needed is a complete modelling of the whole system plus apparatus (S+A) setup, and the dynamics that takes place.
The literature on models for measurement was reviewed by ``ABN'', our collaboration
 with Armen Allahverdyan and Roger Balian, in our 2013 ``Opus Magnum'' \cite{allahverdyan2013understanding}, 
 a paper which we will term  ``Opus'' in the present work.
 A typical early example of measurement models is Hepp's semi-infinite chain of spins $\frac{1}{2}$ which measures the first spin\cite{Hepp1972quantum};
Bell terms it the Coleman-Hepp model\cite{bell2001wave}. 
Gaveau and Schulman consider a ring of such spins and extend the model to measure an atom passing near one of the spins of the ferromagnet.
If the atom is in the excited state, it enhances the phonon coupling of that spin to the lattice, so as to create a critical droplet that flips the overall magnetization\cite{gaveau1990model}.
Another model is the few-degrees-of-freedom setup of an overdamped large oscillator measuring a small one\cite{haake1987overdamped,haake1993classical}.
To employ a Bose-Einstein condensate as a measuring apparatus has been proposed by ABN\cite{allahverdyan2001quantum}.
 There is an whole literature on the puzzling idea that only the environment is needed to describe quantum 
 measurements\cite{zeh1970interpretation,zurek2003decoherence,schlosshauer2005decoherence}.
 
  To back up the popular von Neumann-Wheeler ``theory'' of quantum measurement, put forward in von Neumann's 1932 book on the Hilbert space structure of quantum  mechanics  \cite{von2013mathematische,von2018mathematical}, no working models are known to us, so that
the ensuing ``relative state''\cite{everett1957relative} or ``many worlds interpretation ’''\cite{dewitt1970quantum,dewitt2015many} remains at an intuitive level.
The apparatus is supposed to start and remain in a pure state.
  Our own approach employing Hamiltonians for the measurement dynamics as elaborated in next paragraph, considers the apparatus to start in a metastable thermal state and to end up in a stable one.
 In the  von Neumann-Wheeler philosophy, one would have to slice the initial mixed state in pure components and identify representative ones 
 as ``pure states of the apparatus’'. But these ``states''  interact with each other during the dynamical phase transition
that makes the pointer indicate the outcome, so that the representative sliced ``pure states''  at the final time  
were extremely improbable initially, which makes the connection unnatural.

Progress was made in this millenium, when our ABN collaboration
introduced the ``Curie-Weiss model for quantum measurement''\cite{allahverdyan2003curie}. 
Here for a system S, which is just a spin-$\frac{1}{2}$ that does not evolve in time, the operator $\hat s_z$ is measured by an apparatus A. 
The latter consists of a magnet M and a thermal bath B.
M contains $N\gg1$ spins-$\frac{1}{2}$ and B is a harmonic oscillator bath in a thermal state at temperature $T$.
The model appears to be rich enough to deal with various fundamental issues in quantum measurement.
Many details of the dynamics and subsequently the thermodynamics were worked out in various followup papers \cite{allahverdyan2003quantuma,allahverdyan2005quantum,allahverdyan2005dynamics,allahverdyan2007quantum,allahverdyan2006phase}
and further expanded and summarized in ``Opus'' \cite{allahverdyan2013understanding}.
Lecture notes on the subject were presented \cite{nieuwenhuizen2014lectures}.
A straightforward interpretation for a class of such measurements models was provided \cite{allahverdyan2017sub}.
A paper on teaching the ensuing insights is in preparation \cite{allahverdyan2022teaching}.
A numerical test on a simplified version of the Curie-Weiss model by Donker et al. reproduced nearly all of its properties\cite{donker2018quantum}.

The dynamics of the measurement can be summarized as follows.
In a very small time window after coupling the system S to the apparatus A, there occurs a truncation of the density matrix (erasing 
Schr\"odinger cat terms)  due to first dephasing in the magnet and then decoherence due to the phonon bath. On a longer time scale the 
registration of the measurement takes place because the coupling of S to A allows the magnet to leave its initial paramagnetic state 
and go to the thermodynamically stable state with magnetization upwards or downwards in the $z$-direction, which can then be read off.

The  interpretation of quantum mechanics ensuing from these models is that the density matrix describes our best  knowledge about the 
ensemble of identically prepared systems. 
The truncation of the density matrix (disappearance of Schr\"odinger cat terms) is a dynamical effect, 
while the Born rule follows  in case of an ideal experiment from the dynamical conservation of the tested operator.
A quantum measurement consists of a large set of measurement runs on a large set of identically prepared systems.
Reading off the pointer of the apparatus (the final upward or downward magnetization) 
allows to select the measurement outcomes and to update the predictions for future experimentation.

The insight that quantum mechanics must be only considered in its laboratory context was stressed in particular by 
Bohr, see Max Jammer\cite{jammer1966conceptual}, and is central in the approach of  Auff\`eves and Grangier\cite{auffeves2016contexts,auffeves2020deriving}.
Their contexts-systems-modalities (CSM) approach is complementary to our model based approach.
However, the latter proves rather than postulates the working of the setup and, among others, 
provides specifications for the (model) experiment to be close enough to ideal.

The aim of the present paper is to present Hamiltonians for the measurement of $\hat s_z$ of a higher spin like $l=1,\frac{3}{2},2,\frac{5}{2}$.
To have an unbiased apparatus, M must have a $Z_{2l+1}$ invariance for measuring any of the
eigenvalues of $\hat s_z$, to be denoted as  $s=-l,-l+1,\cdots,l$.
This is achieved by starting from cosines of the spins of M, 
while they allow a simplified connection to low moments of these spins. 
The manifest invariance in the cosine-formulation leads to a linear map between the moments.

In section \ref{sec2} we propose the formulation for the Hamiltonian of M for general spin-$l$.
In section \ref{sec3} we verify that for spin-$\frac{1}{2}$ this leads to the known Curie-Weiss model.
In section \ref{sec4} we consider the thermodynamics of the spin-$1$ situation in detail.
In section \ref{sec5} we investigate the thermodynamics for spins $\frac{3}{2}$, $2$ and $\frac{5}{2}$.
We close with a summary in section \ref{sec7}.

\renewcommand{\thesection}{\arabic{section}}
\section{General spin}
\renewcommand{\theequation}{2.\arabic{equation}}
\setcounter{equation}{0} \setcounter{figure}{0}
\renewcommand{\thesection}{\arabic{section}}
\label{sec2}

\newcommand{\specl}{{\rm spec}_l}
We aim to measure the $z$-component of an arbitrary quantum spin-$l$ with ($l=\frac{1}{2},1,\frac{3}{2},2,\cdots$).
The eigenvalues $s$ of the operator $\hat s_z$ (we indicate operators by a hat) lie in the spectrum\footnote{To simplify the notation, 
we replace the standard notation for spins by $s\to l$ and $s_z\to s$. 
For an angular momentum $ L^2=l(l+1)$, the model also applies to the measurement of  $\hat L_z$
with eigenvalues $m\to s$. We employ units $\hbar=k=1$.}
\BEQ \label{specl}
s \in \specl=\{-l,-l+1,\cdots,l-1,l \} .
\EEQ
The measurement will be performed by employing an apparatus with $N\gg1$ vector spins-$l$ denoted by  {\boldmath{$\hat \sigma^\hi$}},
$i=1,\cdots,N$. . 
They have components $\hat\sigma^\hi_{a}$, $a=x,y,z$, which are coupled to a thermal harmonic oscillator bath; 
for the case $l=\frac{1}{2}$ this was worked out \cite{allahverdyan2003curie,allahverdyan2013understanding}.  
The generalization of such a bath for arbitrary spin-$l$ is straightforward and will be applied to the spin-1 model in future work.

The eigenvalues $\sigma_i$ of each $\hat \sigma^\hi_{z}$ lie also in the spectrum (\ref{specl}).
Since the present work only considers these $z$-components, we can discard the operator nature, and only deal with the eigenvalues,
which are integer or half-integer numbers.

In order to have an unbiased apparatus, the Hamiltonian of the magnet should have maximal symmetry and degenerate minima.
To construct such a functional,  we consider the spin-spin form
\BEQ \label{C2=}
C_2=\frac{1}{N^2}\sum_{i,j=1}^N \cos\frac{2\pi (\sigma_i-\sigma_j) }{2l+1} 
=\left(\frac{1}{N}\sum_{i=1}^N \cos\frac{2\pi\sigma_i }{2l+1}\right)^2
+\left(\frac{1}{N}\sum_{i=1}^N \sin\frac{2\pi\sigma_i }{2l+1} \right)^2  .
\EEQ
The expression in the middle is manifestly invariant under the shift of all $\sigma_i\to\sigma_i+\tilde\sigma$ mod $2l+1$
 for any $\tilde\sigma\in$ spec$_l$.
$C_2$ is non-negative and  lies between 0 for the paramagnet, and 1 for each of the $2l+1$ ferromagnetic states 
where all $\sigma_i$ take one of the values of (\ref{specl}). 
Since the $\sig_i$ in (\ref{C2=}) take the finite number of $2l+1$ values, their cosines and sines can be expressed as polynomials 
of order $2l$ in $\sig_i$, which, summed over $i$,  leads the spin-moments $m_1$, $m_2$, $\cdots$, $m_{2l}$,
where
\BEQ
m_k=\frac{1}{N}\sum_{i=1}^N\sigma_i^k,\qquad (k=1,\cdots,2l) .
\EEQ
Let out of the $N$ spins $\sigma_i$,  a number $N_\sigma=\sum_i\delta_{\sigma_i, \, \sigma}$
 be in state $\sigma$, with $\sigma\in \specl$ and let $x_\sigma =N_\sigma /N$ be their fraction.
The constraint $\sum_{\sigma}N_\sigma =N$ implies $\sum_{\sigma}x_\sigma =1$. The moments read likewise
\BEQ\label{mk=}
m_k=\sum_{\sigma=-l}^l x_\sigma \sigma^k,\quad k=1,\cdots,2l,\qquad m_0=\sum_{\sigma=-l}^l x_\sigma=1.
\EEQ
Inversion of these relations determines the $x_\sigma$ as linear combinations of the $m_k$.
There is no simple general formula for this.  In next sections we work out a number of low-$l$ cases.

\newcommand{\rmA}{{\rm A}}
\newcommand{\rmB}{{\rm B}}
\newcommand{\rmM}{{\rm M}}
\newcommand{\rmS}{{\rm S}}

For the Hamiltonian $H=H_N/N$ we shall follow \cite{allahverdyan2003curie} and adopt the spin-spin and four-spin terms
\BEQ\label{UN=}
H=-\frac{1}{2}J_2C_2-\frac{1}{4}J_4C_2^2 ,
\EEQ
while multi-spin interaction terms like $-\frac{1}{6}J_6C_2^3-\frac{1}{8}J_8C_2^4$ can be added,
but they will not change the overall picture.
In a quantum approach, the $\sigma_i\to\hat \sigma_z^\hi$ and the $m_k \to \hat m_k$ will be operators;
the Hamiltonian of the magnet M will be $\hat H_\rmM=H_N(\hat m_k)$.

The degeneracy of this state is the multinomial coefficient
\BEQ  \label{G=}
\hspace{-3mm}
G={N \choose N_{-l}\,,N_{-l+1}, \cdots  , N_l}
 =\frac{N!}{(N_{-l}) ! (N_{-l+1})!\cdots (N_l)!}  =\frac{N!}{(Nx_{-l}) ! (Nx_{-l+1})!\cdots (Nx_l)!} .
\EEQ
The entropy reads  $S_N=\log G=NS$. 
With the Stirling formula it follows that for large $N$,
\BEQ \label{Sx=}
S=\frac{1}{N}\log G=-\sum_{\sigma=-l}^l  x_\sigma \log x_\sigma .
\EEQ
The thermodynamic  free energy per magnet spin is
\BEQ F=U-TS=\langle H\rangle-T\langle S\rangle.
\EEQ

In order to use the magnet coupled to its bath as an apparatus for a quantum measurement,
a coupling to the system S is needed. In the sector of Hilbert space where the tested quantum operator $\hat s_z$ has the eigenvalue $s$,
the system-apparatus  interaction can likewise be taken as a sum of spin-spin couplings,
\BEQ\label{HSAs} && \hspace{-6mm}
H_\SA^s=N \Is,\qquad  \nn\\&& \hspace{-6mm}
\Is=-\frac{g}{N}\sum_{i=1}^N \cos\frac{2\pi (s-\sigma_i)}{2l+1} 
=-\frac{g}{N}\sum_{i=1}^N\big( \cos\frac{2\pi s}{2l+1} \cos\frac{2\pi \sigma_i}{2l+1} +\sin\frac{2\pi s}{2l+1} \sin\frac{2\pi \sigma_i}{2l+1} \big) ,
\EEQ
where $g$ is the coupling constant.
It will be seen that for given $l$ it can be expressed as a linear combination of the moments $m_1$, $\cdots$, $m_{2l}$.

When the coupling $g$ is turned on, the total free energy per spin is $F_s(m_1,\cdots,m_{2l})=H-TS+I_s$.
At low enough $T$, it has an absolute minimum when nearly all $\sigma_i$ are equal to $s$.
In a measurement setup, one considers quantum dynamics of the system, starting initially in the paramagnetic state
and evolving to this absolute minimum.
In the paramagnet, the spins are randomly oriented, so the fractions $x_\sigma=\frac{1}{2l+1}$ are equal. This leads to the moments
\BEQ
m_k=\frac{1}{2l+1}\sum_{\sigma=-l}^l \sigma^k \qquad  \text{(paramagnet).}
\EEQ
Clearly, the odd moments are zero. The relevant even moments are $m_2=\frac{2}{3}$ for $l=1$;  $m_2=\frac{5}{4}$ for $l=\frac{3}{2}$;
$m_2=2$ and $m_4=\frac{34}{5}$ for $l=2$; and, in the case $l=\frac{5}{2}$ we finally consider,
 $m_2=\frac{35}{12}$ and $m_4=\frac{707}{48}$.

The quantum evolution leads the system from the paramagnet to the lowest free energy  state characterized by $s$,
undergoing a dynamical phase transition and ending with different parameters $m_1,\cdots m_{2l}$.
In a measurement setup, the $2l$ macroscopic order parameters $M_k=Nm_k$ can be read off, and the 
``measured'' value of $s$ can be deduced from them.

The $Z_{2l+1}$ invariance implies that expressions for $C_2$, $U$, $S$, $F$, $\Is $ and $F_s$ 
are invariant under the simultaneous permutations $s\to s'= s+1$ mod $2l+1$ and 
$\sigma_i\to\sigma_i'=\sigma_i+1$ mod $2l+1$ for all $i$. 
For any sequence $\{\sigma_1,\cdots,\sigma_N\}$ the numbers $N_\sigma=\sum_i\delta_{\sigma_i,\,\sigma}$ 
and the fractions $x_\sigma=N_\sigma/N$ are maintained. (An example for $l=1$, $N=4$: 
The sequence $\{1,-1,0,1\}\to \{-1,0,1,-1\}$ has $x_{-1}=x_0=x_0'=x_1' =\frac{1}{4}$ and $x_{1}=x'_{-1}=\frac{1}{2}$. Hence
$m_1=-x_{-1}+x_1=\frac{1}{4}$ and $m_1'=x_{0}-x_1=-x_{-1}'+x_1'=-\frac{1}{4}$, while $m_2'=m_2=\frac{3}{4}$.\label{l=1,N=4})
An equivalent method is to maintain $\sigma$ while introducing $x'_\sigma=x_{\sigma-1\text{  mod }2l+1}$. For $k=1,\cdots 2l$ this gives
\BEQ\label{mk'=}
 m_k'=\sum_{\sigma=-l}^l x_\sigma'\sigma^k=\sum_{\sigma=-l}^l x_\sigma\sigma'{}^k
 = \sum_{\sigma=-l}^l x_{\sigma} (\sigma+1\,\text{mod}\,2l+1 )^k
= \sum_{\sigma=-l}^l x_{\sigma}(\sigma+1)^k +x_{l} [(-l)^k -(l+1)^k] ,  \nn
 \EEQ
which, with $m_0=1$, can be written as the linear map between $s$ and the $m_k$,
 \BEQ \label{mk'=}
 s'=s+1\text{   mod  }2l+1,\qquad 
 m_k'= \sum_{n=0}^k 
 {k\choose n}
 m_{n} + [(-l)^k-(l+1)^k]x_{l} ,\quad (k=1,\cdots,2l).
 \EEQ
When the results for $F_s$ are known for one of the $s$-values,
the results for other $2l$ cases can be obtained from that by applying this map $2l$ times.
Indeed, our starting point with the manifestly invariant cosines in  eqs. (\ref{C2=}) and (\ref{HSAs}) has straightforwardly 
led to this invariance as a map between the moments $m_k$. 
It assures that the apparatus has no bias for measuring any particular $s\in$ spec$_l$ value.

\renewcommand{\thesection}{\arabic{section}}
\section{Recap: The spin-$\frac{1}{2}$ Curie-Weiss model}
\renewcommand{\theequation}{3.\arabic{equation}}
\setcounter{equation}{0} \setcounter{figure}{0}
\renewcommand{\thesection}{\arabic{section}}
\label{sec3}

We will work out the above models for low values of the spin.
We set the stage by considering the spin-$\frac{1}{2}$ situation, a gentle reformulation of the original Curie-Weiss model 
for quantum measurement\cite{allahverdyan2003curie}. 
In units of $\hbar$, the $z$-component of a spin $l=\frac{1}{2}$ has the eigenvalues
\BEQ
s\in \left\{-\frac{1}{2},\frac{1}{2} \right\},
\EEQ
which implies
\BEQ \label{expis12}
\cos\pi s=0,\qquad \sin\pi s=2s.
\EEQ
The magnet has $N$ such spins with each $\sigma_i\in\{-\frac{1}{2},\frac{1}{2}\}$.
According to (\ref{C2=}) we consider the $Z_2$ invariant
\BEQ \label{C2s12}
\hspace{-5mm}
C_2=\frac{1}{N^2}\sum_{i,j=1}^N \cos\pi(\sigma_i-\sigma_j) .
\EEQ
In terms of the moment
\BEQ
m_1=\frac{1}{N}\sum_{i=1}^N \sigma_i.
\EEQ
which lies in the interval $-\frac{1}{2}\le m_1\le\frac{1}{2}$, $C_2$ equals, using (\ref{C2=}) and  (\ref{expis12}) for each $\sigma_i$ and summing, 
\BEQ
C_2=4m_1^2 .
\EEQ
$C_2=0$ for the paramagnetic state $m_1=0$, while
$C_2=1$ when all $\sigma_i$ equal $s=\pm\frac{1}{2}$ and $m_1=s$.

From (\ref{UN=}) the Hamiltonian is taken as pair and quartet interactions,
\BEQ
\label{UN/12=}
H_N=NH,\qquad H=-\frac{J_2}{2}C_2-\frac{J_4}{4}C_2^2=-2J_2m_1^2-4J_4m_1^4 .
\EEQ

With $x_\sigma=N_\sigma/N$ for $\sigma=\pm\frac{1}{2}$ we have from (\ref{mk=})
\BEQ
m_1=\frac{x_{1/2} - x_{-1/2 }}{2} ,
\qquad 
x_{\pm 1/2}=\frac{1\pm 2m_1}{2} .
\EEQ
From (\ref{G=}) and  (\ref{Sx=})  we get the standard result for the entropy at large $N$
\BEQ
S_N=NS,\qquad S= -\frac{1+2m_1}{2}\log\frac{1+2m_1}{2} -\frac{1-2m_1}{2}\log\frac{1-2m_1}{2}.
\EEQ

In order to use the magnet coupled to its bath as an apparatus for a quantum measurement,
a system--apparatus (SA) coupling is needed. It can be chosen as a spin-spin coupling, 
\BEQ\label{HSAs12}
H_\SA^\hs=N\Is ,\qquad \Is
=-\frac{g}{N}\sum_{i=1}^N \cos\pi (s-\sigma_i)
=-\frac{g}{N}\sum_{i=1}^N \sin\pi s\sin\pi\sig_i
=-4gsm_1.
\EEQ
where (\ref{expis12}) was employed also for $\sigma_i$.
The free energy per spin in the $s$-sector, $F_s=H-TS+\Is $, reads
\BEQ \label{Fs1/2}
F_s(m_1)=-2J_2m_1^2 - 4J_4m_1^4 +T\frac{1+2m_1}{2}\log\frac{1+2m_1}{2} +T\frac{1-2m_1}{2}\log\frac{1-2m_1}{2} -4gsm_1.
\EEQ
In accordance with (\ref{mk'=}), it has the invariance $F_{\pm1/2}(\pm m_1)=F_{\mp1/2}(\mp m_1)$ required for an unbiased measurement.
At low $T$, $F_s$ takes its lowest value for $m_1\approx s=\pm\frac{1}{2}$. 
This state is reached near the end of the measurement, after which the apparatus is decoupled from the system by setting $g\to0$.
Eq. (\ref{HSAs12}) shows that an amount of energy $4gNsm_1=2gN|m_1|\approx gN$ has to be added to M for the decoupling.
After a quick relaxation to the nearby thermodynamic minimum of the $g=0$ situation, the pointer, that is, the
macroscopic magnetization $M_1=Nm_1$, can be read off, the sign of which reveals the sought sign of $s$.

The map (\ref{mk'=}) reads here $m_1'=-m_1$, so that the paramagnet $m_1=0$ is its stable point.
This should be because it is the fully random state, which is statistically invariant under permutation.

All of this is a reformulation of the original spin-$\frac{1}{2}$ Curie--Weiss model\cite{allahverdyan2003curie}, 
which involves the notation $s'=2s=\pm1$, $\sigma_i'=2\sigma_i=\pm1$,
so that its $m'\equiv 2m_1$ lies between $-1$ and $+1$.
The couplings $J_{2,4}$ in (\ref{UN=}) and $g$ in (\ref{HSAs12})  keep their values;
for example, the interaction term $-4gsm$ in (\ref{HSAs12}) lies for $s=\pm\frac{1}{2}$ and $\frac{1}{2} \le m_1\le\frac{1}{2}$ between $-g$ and $g$,
as does $-gs'm'$ in \cite{allahverdyan2003curie}.
This occurs by construction, since in the definitions (\ref{C2s12}) and  (\ref{HSAs12}),
one has to adjust the arguments of the cosines, not their values.

\renewcommand{\thesection}{\arabic{section}}
\section{The spin-1 Curie-Weiss model} 
\renewcommand{\theequation}{4.\arabic{equation}}
\setcounter{equation}{0} \setcounter{figure}{0}
\renewcommand{\thesection}{\arabic{section}}
\label{sec4}

We now work out similar steps in the model of section \ref{sec2} for spin 1 and analyze the thermodynamics.

\subsection{Formulation of the model}

A spin-1 has discrete $z$-components $s=0,\pm1$.
Since $s^{2k+1}=s$ and $s^{2k+2}=s^2$ for $k\ge 1$, the 3 values of the cosine and sine
can be expressed as quadratic or linear polynomials in $s$,
\BEQ \label{exps2s}
\cos\frac{2\pi s}{3}=1-\frac{3}{2}s^2 ,
\quad
\sin\frac{2\pi s}{3}=  \frac{\sqrt{3}}{2}s . 
\EEQ
Using this with $s\to\sigma_i$ and summing over $i$ leads to  introduce the moments
\BEQ \label{m12def}
m_1=\nu\sum_i \sigma_i,\qquad m_2=\nu\sum_i \sigma_i^2, \qquad \nu\equiv\frac{1}{N} . 
\EEQ
Let $N_\sigma$ denote the number of spins with $\sigma_i =\sigma$ for $\sigma=0,\pm1$.
In terms of the fractions $x_\sigma=N_\sigma/N$ it holds that
\BEQ
m_k=\sum_{\sigma=-l}^l x_\sigma \sigma^k,\qquad  m_0=x_{-1}+x_0+x_1=1,\quad 
m_1=-x_{-1}+x_1,\qquad m_2=x_{-1}+x_1 .
\EEQ
Their inversion reads
 \BEQ\label{xi=}
 x_0=1-m_2,\qquad x_{\pm 1}=\frac{m_2\pm m_1}{2} .
\EEQ
For these to be nonnegative, the physical values are limited to
\BEQ
-m_2\le m_1\le  m_2,\qquad 0\le m_2\le 1,
\EEQ
The actual values of $m_{1,2}$ are found as follows from (\ref{m12def}).
When all $\sigi=0$, $m_1=m_2=0$. 
When one $\sigi=\pm 1$, $m_2=\nu$ and $m_1=\pm\nu$, where $\nu\equiv1/N$; when two of the $\sig_i$ are 
$\pm 1$, $m_2=2\nu$ and $m_1=\pm2\nu$ or $0$; when 3 are $\pm1$, $m_2=3\nu$ and $m_1=\pm 3\nu$ or $\pm \nu$, and so on.
Thus $m_2$ ranges from 0 to 1 with steps of $\nu$,  while $m_1$ ranges from $-m_2$ to $m_2$ with steps of $2\nu$.

To construct the energy, we consider the $Z_3$ invariant of the spins of the magnet
\BEQ
\hspace{-7mm}
C_2=\frac{1}{N^2}\sum_{i,j=1}^N
\cos\frac{2\pi}{3} (\sigma_i-\sigma_j).
\EEQ
Expanding the cosine, employing (\ref{exps2s}) for the $\sigma_i$ and summing yields a polynomial in the moments $m_{1,2}$,
\BEQ\label{Cl=1}
C_2=(1-\frac{3}{2}m_2)^2+\frac{3}{4}m_1^2 . 
\EEQ
For the Hamiltonian we take as in  (\ref{UN=})
\BEQ
H_N=NH,\qquad H=-\frac{1}{2}J_2C_2-\frac{1}{4}J_4C_2^2 .
\EEQ

The degeneracy of states characterized by $m_{1,2}$ is the multinomial
\BEQ
G={N \choose  N_\meen ,\, N_0, \,N_1 } =\frac{N!}{(Nx_\meen)!\, (Nx_0)!\, (Nx_1)!},
\EEQ
Eq. (\ref{Sx=}) yields the explicit result for the entropy per spin $S=S_N/N=(\log G)/N$ at large $N$,
\BEQ
S=-(1-m_2)\log(1-m_2)    -\frac{m_2+m_1}{2}\log\frac{m_2+m_1}{2}  -\frac{m_2-m_1}{2}\log\frac{m_2-m_1}{2}  . 
\EEQ

The $Z_3$ symmetry of these quantities can be expressed by considering the permutation  (\ref{mk'=}), 
\BEQ \label{m12p}
m_1'=1-\frac{1}{2}m_1-\frac{3}{2}m_2,   \qquad
m_2'=1+\frac{1}{2}m_1-\frac{1}{2} m_2 ,
\EEQ
as can be verified in the special case in the second alinea after eq. (10).
It follows that $C_2'=(1-\frac{3}{2}m_2')^2+\frac{3}{4}m_1'{}^2=C_2$ is unchanged.
As expected, the weights (\ref{xi=}) are permuted,
\BEQ
x_{-1}'=\frac{m_2'-m_1'}{2}=x_0,
\quad x_0'=1-m_2'=x_{1}, \quad x_1'=\frac{m_2'+m_1'}{2}=x_{-1}, 
\EEQ
so that $U$, $S$ and $F$ are invariant,  as required for unbiased measurement, and implying that the minima of $F$
are degenerate, see figure 1. (Due to (\ref{G=}), $S_N=\log G$ is invariant at any finite $N$).
Making the shift $s\to s'=s+1$ mod 3 a second time, or the inverse shift $s\to s'=s-1$ mod 3,  leads  (\ref{m12p}) to
\BEQ \label{m12d}
m_1''= -1-\frac{1}{2}  m_1+ \frac{3}{2} m_2  ,\qquad
m_2''=1 -\frac{1}{2} m_1 - \frac{1}{2} m_2.
\EEQ
Inserting (\ref{m12p}) in the right hand side of $m_{1,2}''$ leads to  $m_{1,2}'''=m_{1,2}$,  the identity map,  as it should.

The thermodynamic free energy is
\BEQ\label{Fl=1}
F_N=NF,\qquad F=-\frac{1}{2}J_2C_2-\frac{1}{4}J_4C_2^2-TS .
\EEQ
The ferromagnetic states $m_1=m_2=0$ and $m_1=\pm1$, $m_2=1$ have $C_2=1$ and $S=0$.
The paramagnet ($ m_1=0$, $m_2=\frac{2}{3}$) has energy zero and maximal entropy per spin, $S=\log 3$.

For $T$ low enough, one can use the model as a  measuring apparatus that starts in the metastable paramagnetic state and
ends up in one of the three degenerate stable states.
To measure the $z$-component  $s=0,\pm1$ of a spin-1,  we assume that the tested spin S has a spin-spin coupling 
with all spins of the apparatus. For the SA coupling (\ref{HSAs}) we get, using (\ref{exps2s}) for $s$ and the $\sigma_i$,
\BEQ \label{HSA1}
H_{\SA}^\hs=N\Is,\qquad \Is=-\frac{g}{N}\sum_{i=1}^N\cos\frac{2\pi}{3}(s-\sigma_i)
=-g\big [(1-\frac{3}{2}s^2)(1-\frac{3}{2}m_2)+\frac{3}{4}s \, m_1\big] .
\EEQ
It is invariant for $s\to s'=s+1$ mod 3 and likewise for the $\sigi$, the latter being equivalent to $m_{1,2}\to m_{1,2}'$ 
as given in (\ref{m12p}), which, with the invariance of $H_N$ and $S_N$,  assures absence of bias in the measurement.

\subsection{The paramagnetic state}

The paramagnetic state has $m_1=0$, $m_2=\frac{2}{3}$.  It is invariant under the map (\ref{m12p}), 
as expected, since it refers to the completely random state.
One may verify the total weight of this state for large $N$ in the Stirling approximation, which leads to 
small Gaussian deviations  $\delta m_1=m_1-0$ and $\delta m_2=m_2-\frac{2}{3}$,
\BEQ
\frac{1}{3^N}\sum_{m_2=0}^1\sum_{m_1=-m_2}^{m_2}G
\approx 
\frac{1}{3^N}\int_\minfty^\infty\frac{\d \delta m_2}{\nu}\int_\minfty^\infty\frac{\d \delta m_1}{2\nu}
\,\, \frac{3^{N+3/2}}{2\pi N}  e^{-(3N/4)\delta m_1^2 - (9N/4) \delta m_2^2}=1.
\EEQ
 Expansion brings likewise $C_2\approx \frac{3}{4}(\delta m_1^2+3\delta m_2^2)$, which yields 
\BEQ 
F=-T\log 3+\frac{3}{4}(T-\frac{J_2}{2})(\delta m_1^2+3\delta m_2^2)+{\cal O}(\delta m_1^4,\delta m_1^2\delta m_2^2,\delta m_2^4) .
\EEQ
At high $T$ the paramagnetic state is the only stable state. At lower $T$
it remains locally stable for $T>\frac{1}{2} J_2$; in case $J_2\le 0$ it is locally stable at all $T$. 
In the measurement setup, this local stability is required to let the apparatus lie in the metastable paramagnetic state
(``ready state'') until the measurement is started.

\subsection{The equilibrium states of the magnet}

The free energy $F(m_1,m_2)$ is given by (\ref{Fl=1}) and (\ref{Cl=1}). For the case $J_2=0$, $J_4=1$ and $T=0.4$, it is depicted in figure 1.
   It has 3 minima, of which one occurs at $m_1=0$ and small $m_2$. At $m_1=0$ one has
\BEQ
F=-\frac{J_2}{2}(1-\frac{3}{2}m_2)^2-\frac{J_4}{4}(1-\frac{3}{2}m_2)^4 +\, T (1-m_2)\log(1-m_2) +T m_2\log\frac{m_2}{2} .
\EEQ
Its mean field equation reads
\BEQ \label{m2mf}
m_2=\frac{2}{e^{h/T}+2} 
,\qquad
h\equiv3J_2(1-\frac{3}{2}m_2)+3J_4(1-\frac{3}{2}m_2)^3 .
\EEQ

\myskip{ }{
\begin{figure}
\label{fig1}
\centerline{ \includegraphics[width=8cm]{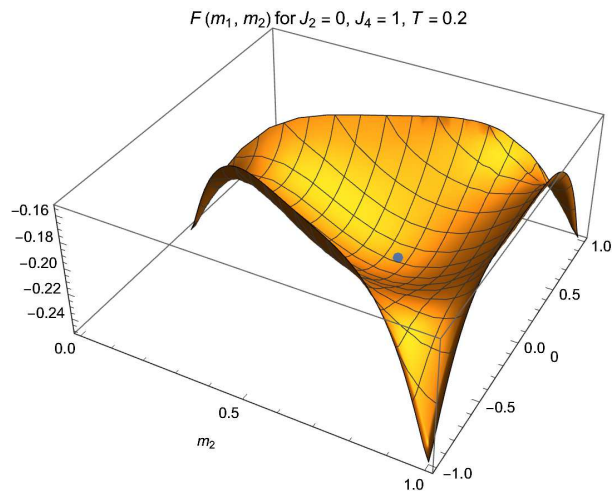}} 
\caption{The free energy $F$$(m_1,m_2)$ of the spin-1 magnet with $J_2=0$, $J_4=1$ at  $T=0.2$ below $T_c=0.228165$.
The physical parameter range is $|m_1|\le m_2\le1$.
The paramagnetic state at ($0,\frac{2}{3}$), indicated by the dot,  is metastable with $F=-0.219722$; 
the 3 minima near the edges are degenerate and stable.
The left one is located at $m_1^\ast=0$, $m_2^\ast=0.00114849$, where thermal effects make
$m_2^\ast>0$ and $F=-0.2502251$ lie below the edge value $-0.25$.
Two other minima lie at the symmetry points $m_1'=1-\frac{3}{2}m_2^\ast$, $m_2'= 1-\frac{1}{2} m_2^\ast$ and 
$m_1''= -1+\frac{3}{2} m_2^\ast$, $m_2''=1 - \frac{1}{2} m_2^\ast$.
In this setting, the magnet can be employed for quantum measurement. 
In the final state one reads off $M_1=Nm_1$, which is close to $0$ or $\pm N$, and $M_2=Nm_2$, 
which is close to 0 or $N$, well separated from the initial paramagnetic values $M_1=0$,  $M_2=\frac{2}{3}N$.
}
\end{figure}
}

\noindent 
The paramagnetic state having $m_2=\frac{2}{3}$ and $h=0$ is the only stable state at high $T$.
In case $J_2=0$, there appears a metastable (ms) state $m_1=0$, $m_2>0$  when $\p_{m_2}F=\p_{m_2}^2F=0$ develops a solution at 
\BEQ
T_{\rm ms}=0.328257\,J_4,    \qquad m_2^{\rm ms}=0.0634132. 
\EEQ
Below the critical temperature $T_c$ this state attains the absolute minimum of the free energy,
\BEQ
T_c=0.228165\,J_4,\qquad \, \, m_2^c= 0.00304442.
\EEQ
Let for general $T<T_c$, $F$  have an absolute minimum at $m_1^\ast=0$ and small $m_2^\ast$; for $T=0.2J_4$ as in figure 1,
$m_2^\ast=0.00114849$.
The $Z_3$ symmetry ensures that this minimum is degenerate with the pair of minima at the symmetry points 
$(m_1',m_2')=(1-\frac{3}{2}m_2^\ast, 1-\frac{1}{2} m_2^\ast)$ and  $(m_1'',m_2'')= (-1+\frac{3}{2} m_2^\ast,1 - \frac{1}{2} m_2^\ast)$.
With $m_2^\ast$ being small,  the minima lie close to the edge values $(m_1,m_2)=(0,0)$, $(1,1)$  and $(-1,1)$, respectively,
where the magnet is polarized with nearly all $\sigma_i$ equal $0,1$ and $-1$, respectively.

When $J_2\neq 0$, it may be negative, but $\frac{1}{2} J_2>-\frac{1}{4}J_4+T\log 3-Tm_2^\ast+{\cal O}(m_2^{\ast\,2})$ is needed
for the minimum at $m_2^\ast$ to have a lower free energy than the paramagnet and thus be the absolute minimum.

\subsection{The thermodynamic equilibrium state of M coupled to S}

The total free energy per particle $F_s=U-TS+\Is$ has an absolute minimum for each $s$, which is most easily analyzed for low $T$.
For $s=0$ it is optimal to have $m_1=0$, $m_2\approx 0$, which occurs when nearly all $\sigma_i$ are $0$;
for $s=\pm1$ it is optimal to have $m_1\approx s$, $m_2\approx 1$, which occurs at low $T$ when nearly all $\sigma_i$ are equal to $s$.
This correlation between the apparatus spins $\sigma_i$ and tested spin $s$ allows to employ the setup as an apparatus that measures $s$
by reading off the macroscopic order parameters $M_{1,2}=Nm_{1,2}$ of the magnet.
Hereto one sets $g$ from 0 to a large enough positive value at an initial time $t_i$ of the measurement and puts it back to zero near the final time $t_f$.
In the first stage, the magnet goes at given $s$ to the state with lowest $F_s$; after cutting $g$, there occurs a small rearrangement
to the nearby stable state of $F$. Then the macroscopic order parameters $M_{1,2}$ can be read off, which determine $s$. 

At large enough  $g$ and proper low $T$, $F_s(m_1,m_2)$ has one absolute minimum for each $s$;
see figure 2 for the case $s=1$. It suffices to know $F_s$ for one of the cases, say $s=1$.
The profiles for $s=-1,0$ read, in the notations of (\ref{m12p}) and (\ref{m12d}),
$F_{-1}(m_1,m_2)=F_1(m_1',m_2')$ and $F_0(m_1,m_2)=F_1(m_1'',m_2'')$.

The free energy at $m_1=0$ is relevant in the case $s=0$. With  the SA interaction $I_0 $ included, it reads
\BEQ  \hspace{-6mm}
F_0=-\frac{J_2}{2}(1-\frac{3}{2}m_2)^2-\frac{J_4}{4}(1-\frac{3}{2}m_2)^4
+T (1-m_2)\log(1-m_2) +T m_2\log\frac{m_2}{2} -g(1-\frac{3}{2}m_2)  .
\EEQ
Its mean field equation (\ref{m2mf}) now includes $g$,
\BEQ &&
m_2=\frac{2}{e^{h/T}+2} 
,\qquad 
h=3J_2(1-\frac{3}{2}m_2)+3J_4(1-\frac{3}{2}m_2)^3+\frac{3}{2}g .
\EEQ
The paramagnet $m_2=\frac{2}{3}$ is not a solution at $g\neq0$.
When $J_2=0$ and $T=0.4J_4$, a coupling $g>g_c= 0.170642\,J_4$ is needed to suppress the barrier around $m_2=0.4352046$
between the paramagnetic and $F_0$ state, so that the ferromagnetic pointer state can be reached dynamically by ``sliding off the hill''.

For small $T$, which can be used when $J_2$ is small or negative but $T>\frac{1}{2} J_2$,   $m_2$ is exponentially small, 
\BEQ
m_2\approx 2e^{-3( J_2 +  J_4+g)/2T} .
 \EEQ
The free energy, equal to
\BEQ
F_0=-\frac{1}{2}J_2-\frac{1}{4}J_4-Tm_2+{\cal O}(m_2^2), 
\EEQ
lies slightly below the corner value at $m_1=m_2=0$ and 
well below the paramagnetic  $F=-T\log 3$. The free energy $F_s$ for the case $s=1$ is plotted fig. 1 as function of $m_{1,2}$.
For $m_1=0$ it is plotted as function of $m_2$ in figure 3, both for $g=0$ and $g\neq0$.

The stability of a extremal state with $m_1^\ast=0$ and finite $m_2^\ast$  is set by
\BEQ 
\p_{m_1}^2F= \frac{T}{m_2^\ast}-\frac{3}{4} J_2 - \frac{3}{4}J_4 (1 -\frac{ 3}{2} m_2^\ast)^2, \quad
\p_{m_2}^2F =
\frac{T}{m_2^\ast(1 - m_2^\ast)}
-\frac{9}{4} J_2 - \frac{27}{4}J_4 (1 -\frac{ 3}{2} m_2^\ast)^2  , 
  \EEQ
  while $\p_{m_1}\p_{m_2}F =0$.
For small $T$ one has $m_2^\ast\ll 1$, so these are approximately equal to $T/m_2^\ast$, making this point is stable. 
There are two related stable points: The minimum of $F$ at $m_1^\ast=0$, $m_2^\ast>0$ is degenerate with   
 $m_1^\ast{}'=\pm(1-\frac{3}{2}m_2^\ast)$, $ m_2^\ast{}'=1-\frac{1}{2}m_2^\ast$. In all cases,  $m_1^\ast\approx s=0,\pm1$ and
 $m_2^\ast\approx s^2=0,1$.

\begin{figure}
\label{fig2}
\centerline{ \includegraphics[width=8cm]{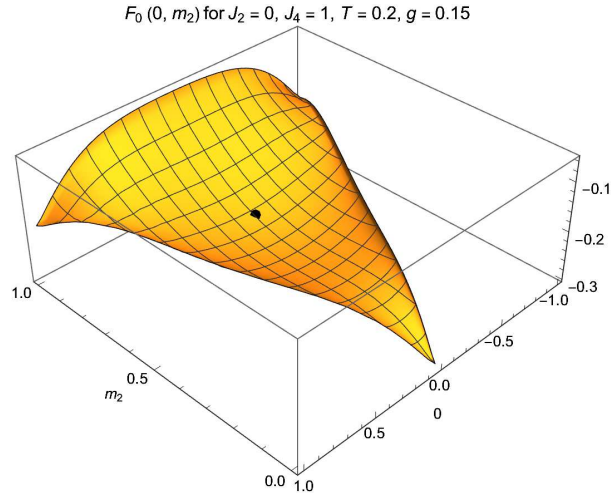}} 
\caption{The free energy $F_s(m_1,m_2)$ of the spin-1 magnet with $J_2=0$, $J_4=1$ 
at  $T=0.2$ coupled to the spin-1 with strength $g=0.15$ in the sector $s=0$. 
The coupling acts as a magnetic field, leading the magnet from its initial paramagnetic state at ($0,\frac{2}{3}$), 
indicated by the dot,  to the absolute minimum of $F_0$ at $m_1^\ast=0$ and small $m_2^\ast$. For $s=\pm1$,  $F_s$ is related by the maps (\ref{m12p}), (\ref{m12d}), viz. 
$F_{1}(m_1,m_2)=F_0(m_1',m_2')$ and $F_{-1}(m_1,m_2)=F_0(m_1'',m_2'')$.
}
\end{figure}

\begin{figure}
\label{fig3}
\centerline{ \includegraphics[width=8cm]{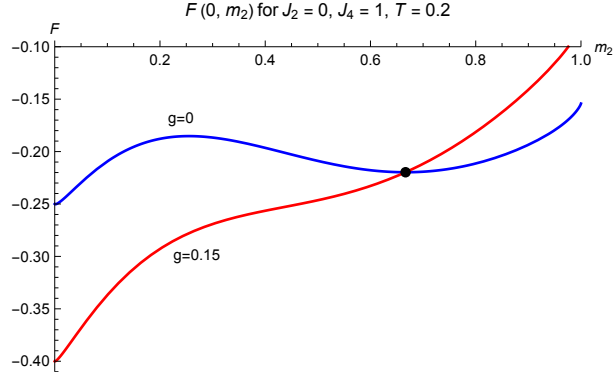}} 
\caption{The free energy $F_s(m_1=0,m_2)$ of the spin 1 magnet with parameters as in figs. 1 and 2:  $J_2=0$, $J_4=1$ 
at  $T=0.2$, (not) coupled to a spin $1$ with strength $g=0.15$ in the sector $s=0$. 
The coupling acts as a magnetic field, leading the magnet from its initial paramagnetic state 
indicated by the dot,  to the absolute minimum at $m_1^\ast=0$ and small $m_2^\ast$.
}
\end{figure}

\subsection{Spin 1 effectively behaving as spin $\frac{1}{2}$}

If $m_2=1$, the value $x_0=0$ shows that $\sig=0$ states are empty, so that only the $\sig=\pm1$ states participate,
effectively a spin $\frac{1}{2}$ system. This can be achieved by a strong repulsive magnetic field in the 0-direction,  
expressed by the Hamiltonian $\Delta H_N=H_0\sum_{n=1}^N (1-\sig_n^2)=NH_0x_0$ with $H_0\gg J_{2,4}$.

\subsection{An apparatus that measures only two values of $s_z$ of a spin 1}

Suppose that we couple the $l=1$ spin S to an apparatus with spins $\sigma_i=\pm \frac{1}{2}$, which have $|m_1|\le\frac{1}{2}$ and $m_2=\frac{1}{4}$.
Let the interaction Hamiltonian not be set by (\ref{HSA1}) but by
\BEQ \label{HSA1i}
\Is=-\frac{g}{N}\sum_{i=1}^N\cos\frac{2\pi}{3}(s-2\sigma_i)
=\frac{g}{2}(1-\frac{3}{2}s^2)-\frac{3}{2} \,gs m_1 .
\EEQ
In each sector $s=0,\pm1$, the first term is a constant that can be dropped. 
For $s=\pm 1$ registration takes place as in the spin-$\frac{1}{2}$ CW model of section 2, where the sign of the final total magnetization 
$M_1=Nm_1$ is set by the sign of $s$. 
In the sector $s=0$ there is no coupling between system and apparatus, 
hence no dynamics takes place: the apparatus does not act;
not even a truncation of the density matrix occurs.

Setting $s\to -\delta_{s,-1}+\delta_{s,0}$ in eq. (\ref{HSA1i}) brings a model for measuring only the $s=-1$ and $s=0$ values of $s_z=0,\pm1$,
while $s\to -\delta_{ s,0}+\delta_{s,1}$ leads to a model for measuring $s=0$ and $s=1$.

\renewcommand{\thesection}{\arabic{section}}
\section{Higher spin models}
\renewcommand{\theequation}{5.\arabic{equation}}
\setcounter{equation}{0} \setcounter{figure}{0}
\renewcommand{\thesection}{\arabic{section}}
\label{sec5}

Proceeding in a similar way as for spin $\frac{1}{2}$ and $1$, we consider the cases of spin $l=\frac{3}{2}$, $2$ and $\frac{5}{2}$.

\subsection{Spin 3/2}

The $z$-component of a spin $l=\frac{3}{2}$ can take the values
\BEQ\label{specl32}
s\in \left \{-\frac{3}{2},-\frac{1}{2},\frac{1}{2},\frac{3}{2} \right \}.
\EEQ
This implies
\BEQ
e^{\pi i s/2} =\left \{ \frac{-1-i}{\sqt}, \frac{1-i}{\sqt}, \frac{1+i}{\sqt},\frac{1-i}{\sqt} \right\},
\EEQ
which may be expressed by the at most cubic polynomials
\BEQ
\cos\frac{\pi s}{2}=\frac{5-4s^2}{4\sqt} ,\quad \sin\frac{\pi s}{2}=\frac{13s-4s^3}{6\sqt} .
\EEQ

The magnet has $N$ such spins $\sigma_i$. 
We consider the $Z_4$ invariant
\BEQ
\hspace{-5mm}
C_2=\frac{1}{N^2}\sum_{i,j=1}^N \cos\frac{\pi}{2} (\sigma_i-\sigma_j) .
\EEQ
Expressed in the magnetic moments takes the form
\BEQ
C_2=\frac{(5-4m_2)^2}{32}+\frac{( 13 m_1-4m_3)^2}{72} .
\EEQ
with the standard definitions
\BEQ
m_k=\frac{1}{N}\sum_{i=1}^N \sigma_i^k,\qquad k=1,2,3.
\EEQ
$C_2=1$ when all $\sigma_i$ equal any of the $s$ values,  in which $m_k=s^k$, as it should.

In the paramagnet one has random $\sigma_i$, each taking one of the $s$ values with probability $\frac{1}{4}$,
\BEQ \label{PMs32}
m_k\to \frac{1}{4}\sum_{\sigma=-3/2}^{3/2} \sigma^k,\qquad m_1,m_3\to 0,\quad  m_2\to\frac{5}{4}. 
\EEQ

The multinomial
\BEQ
G={ N \choose  N_{-\frac{3}{2}} ,\, N_{-\frac{1}{2}}, \,N_{\frac{1}{2}}, \,N_{\frac{3}{2}} } ,\qquad N_\sig=Nx_\sig,
\EEQ
leads for large $N$  to the entropy per particle
\BEQ
S= - x_{-\frac{3}{2}}\log x_{-\frac{3}{2}}-x_{-\frac{1}{2}}\log x_{-\frac{1}{2}}
-x_{\frac{1}{2}} \log x_{\frac{1}{2}}-x_{\frac{3}{2}}\log x_{\frac{3}{2}} .
\EEQ
The spin moments are
\BEQ
m_k=\sum_{\sigma=-l}^l x_\sigma \sigma^k 
=\frac{1}{2^k}\left[x_{1/2}+(-1)^kx_{-1/2}\right]+ \frac{3^k}{2^k}  \left[ x_{3/2}+(-1)^kx_{-3/2}\right].
\EEQ
Inversion of the $k=1,2,3$ expressions brings
\BEQ
 x_{\pm\frac{1}{2}}=\frac{ 9 \pm 18 m_1 - 4 m_2 \mp 8 m_3}{16} , 
\qquad\qquad 
x_{\pm\frac{3}{2}}=\frac{-3 \mp 2 m_1 + 12 m_2 \pm 8 m_3}{48} .
\EEQ
They must all be nonnegative, which confines the allowable $m_k$. This implies
$|2m_1-8m_3|\le 12(m_2-\frac{1}{4})$ and $|18m_1-6m_3|\le 4(\frac{9}{4}-m_2)$.
The combinations $x_{-3/2}+x_{3/2} \ge 0$ and $x_{-1/2}+x_{1/2} \ge 0$ impose $\frac{1}{4}\le m_2\le \frac{9}{4}$,
in accordance with its definition $m_2=\sum_{\sigma=-3/2}^{3/2}x_\sigma \sigma^2$. 
The combinations $3x_{-3/2}+x_{-1/2}$ and  $3x_{3/2}+x_{1/2}$ yield $|m_1|\le \frac{1}{2} m_2+\frac{3}{8}$,
while $27x_{-3/2}+x_{-1/2}$ and  $27x_{3/2}+x_{1/2}$ yield $|m_3| \le \frac{13}{8}m_2-\frac{9}{32}$.

In $F_s=U-TS+\Is$ the coupling to S is chosen as in (\ref{HSAs}), 
\BEQ 
\Is=-\frac{g}{N}\sum_{i=1}^N \cos\frac{\pi (s-\sigma_i)}{2}=
-g \left[\frac{(5-4s^2)(5-4m_2)}{32}+\frac{( 13 s-4s^3)( 13 m_1-4m_3)}{72} \right] .
\EEQ
Again it leads to the lowest value $H_\SA=-gN$, when $m_k=s^k$ for all $\sigma_i=s$ for all four $s$-values in the spectrum (\ref{specl32}).
Due to thermal effects, the optimal $m_k$ will slightly deviate from these values.

The permutation $s \to  s'=s+1\text { mod } 4$ and   $\sigma_i \to  \sigma_i'=\sigma_i+1\text { mod } 4$
leads to (\ref{mk'=}) in  the form
\BEQ \hspace{-4mm}
\{m_1',m_2',m_3'\}=\{\frac{5}{4} + \frac{7}{6} m_1- m_2 - \frac{2}{3} m_3,  
\frac{5}{4} + \frac{13}{6} m_1 - \frac{2}{3} m_3,
\frac{35}{16} + \frac{91}{24} m_1 -\frac{7}{4}m_2 - \frac{13}{6} m_3\}. 
  \EEQ
This leaves $S$, $C_2$, and $\Is $ and hence $U$, $F$ and $F_s$, invariant.

The paramagnetic state (\ref{PMs32}) is invariant under the map, as it should for the completely random state.

\subsection{Spin 2}

The $z$-component of a spin $2$ takes one of the values
\BEQ
s \in \{-2,-1,0,1,2\}.
\EEQ
It is handy to define
\BEQ 
&& {\rm co}(m_2,m_4)=1 -\frac{ 75 - 17 \sqf  }{48} m_2  +\frac{ 5 (3 - \sqf )\, }{48}m_4 ,\quad  \nn\\
&& {\rm si }(m_1,m_3) =\frac{m_1}{24} \sqrt{2 (325 + 31 \sqf)}  -\frac{m_3}{24}\sqrt{ 10  (5 - \sqf) }   ,
 \EEQ
which arise from the properties 
 \BEQ  
 \cos\frac{2\pi s}{5} =  {\rm co}(s^2,s^4), \qquad \sin\frac{2\pi s}{5} = \,{\rm si }(s,s^3)  .
\EEQ
From eq. (\ref{C2=}) it follows that
\BEQ \label{C2l2}
C_2={\rm co} ^2(m_2,m_4)+{\rm si }^2(m_1,m_3) .
\EEQ

In the paramagnet one has random $\sigma_i$, so that
\BEQ
m_k\to \frac{1}{5}\sum_{s=-2}^2 s^k
\EEQ
which amounts to $m_1=m_3=0$, $m_2=2$, $m_4=34/5$, confirming that $C_2$ vanishes.
The moments read
\BEQ
m_k=(-2)^kx_{-2}+(-1)^kx_{-1}+x_1+2^kx_2 \qquad (k=1,2,3,4).
\EEQ
 The $x_k$ follow as
\BEQ \hspace{-3mm}
  x_0=\frac{4 - 5 m_2 + m_4}{4} ,  \quad   x_{\pm1} =\frac{\pm 4 m_1 + 4 m_2 \mp m_3 - m_4}{6},   
  \quad   x_{\pm 2} =\frac{ \mp\, 2 m_1 - m_2 \pm 2 m_3 + m_4}{24} ,
\EEQ
and fix the entropy by (\ref{Sx=}).
The ferromagnetic state $m_k=s ^k$ indeed has $x_{\sigma}=\delta_{\sigma,s}$ and $S=0$.
The possible values of $m_k$ follow from $m_k=(1/N)\sum_i\sig_i^k$, which make the $x_\sig$ nonnegative.

The degeneracy (\ref{G=}) leads for large $N$ to the entropy
\BEQ \label{SNs2}
S=\frac{S_N}{N}=-x_{-2}\log x_{-2}-x_\meen \log x_\meen-x_0\log x_0-x_1\log x_1-x_2\log x_2.
\EEQ

The SA coupling (\ref{HSAs}) reads explicitly
\BEQ \label{hsl2}
 H_\SA^\hs=N\Is,\qquad 
\Is=-g \big [ {\rm co} (s^2,s^4) {\rm co} (m_2,m_4)+ {\rm si }(s,s^3) {\rm si }(m_1,m_3) \big]
\EEQ

The map (\ref{mk'=}) takes the form
\BEQ && \hspace{-14mm}
m_1'=1+\frac{17 {m_1}}{12}+\frac{5 {m_2}}{24}-\frac{5 {m_3}}{12}-\frac{5 {m_4}}{24},\qquad \,\,
m_2'=1+\frac{29 {m_1}}{12}+\frac{29 {m_2}}{24}-\frac{5 {m_3}}{12}-\frac{5 {m_4}}{24}  , \nn\\&& \hspace{-14mm}
m_3'=1+\frac{71 {m_1}}{12}+\frac{107 {m_2}}{24}-\frac{23 {m_3}}{12}-\frac{35 {m_4}}{24}, \hspace{2mm}
m_4'=1+\frac{113 {m_1}}{12}+\frac{209 {m_2}}{24}-\frac{17 {m_3}}{12}-\frac{41 {m_4}}{24}.
\EEQ
It leaves $C_2$ of eq. (\ref{C2l2}) invariant as well as (\ref{SNs2}),  and, with $s\to s'=s+1$ mod 5, also the $\Is $ of eq. (\ref{hsl2}).
The paramagnet is the stable point of this map.

\subsection{Spin $\frac{5}{2}$}

Finally we consider $l=\frac{5}{2}$, where
\BEQ
s\in \left\{-\frac{5}{2}, -\frac{3}{2},-\frac{1}{2},\frac{1}{2},\frac{3}{2} ,\frac{5}{2} \right\}.
\EEQ
Here we define
\BEQ 
{\rm co}(m_2,m_4)=\frac{1}{\sqrt{3}}\left( \frac{441}{256} - \frac{29}{32} m_2+\frac{ m_4}{16} \right), \quad  
 {\rm si }(m_1,m_3,m_5) =
  \frac{2009 m_1}{1920} - \frac{3 m_3}{16} + \frac{m_5}{120} .
  \EEQ
which are introduced to satisfy
\BEQ
\cos\frac{\pi s}{3}={\rm co}(s^2,s^4), \qquad
\sin\frac{\pi s}{3}=  {\rm si }(s,s^3,s^5).
\EEQ
This leads to 
\BEQ \label{C2l52}
C_2 = {\rm co} ^2(m_2,m_4)+ {\rm si }^2(m_1,m_3,m_5) ,\qquad U=-\frac{J_2}{2}C_2-\frac{J_4}{4}C_2^2.
\EEQ
and, from (\ref{HSAs}), 
\BEQ  \label{hsl52}
 \Is = - g  \big [ {\rm co} (s^2,s^4) \,  {\rm co} (m_2,m_4)+ {\rm si }(s,s^3,s^5) \, {\rm si }(m_1,m_3,m_5) \big]
\EEQ
The entropy per spin reads at large $N$
\BEQ
S=-x_{-\frac{5}{2}}\log x_{-\frac{5}{2}} - x_{-\frac{3}{2}}\log x_{-\frac{3}{2}}-x_{-\frac{1}{2}}\log x_{-\frac{1}{2}}
-x_{\frac{1}{2}} \log x_{\frac{1}{2}}-x_{\frac{3}{2}}\log x_{\frac{3}{2}} -x_{\frac{5}{2}}\log x_{\frac{5}{2}}.
\EEQ

The weights take the form
\BEQ 
x_{\pm1/2}&=& \,\,\,\frac{75}{128}\pm \frac{75 {m_1}}{64} - \frac{17 {m_2}}{48}\mp\frac{17 {m_3}}{24}+\frac{{m_4}}{24}\pm\frac{{m_5}}{12} , \nn\\
x_{\pm 3/2}&=& \! -\frac{25}{256}\mp\frac{25 {m_1}}{384}+\frac{13 {m_2}}{32}\pm\frac{13 {m_3}}{48}-\frac{{m_4}}{16}\mp \frac{{m_5}}{24},   \nn \\
x_{\pm5/2}&=&\,\, \, \frac{3}{256} \, \pm \,\frac{3 {m_1}}{640} \, \, -\frac{5 {m_2}}{96} \, \,\,\mp \, \, \, \frac{{m_3}}{48} 
\,\, +\frac{{m_4}}{48} \pm \frac{{m_5}}{120} .
\EEQ
They must be nonnegative, which sets the physical ranges of the $m_k$, next to $|m_{k}|\le (5/2)^{k}$  for $k=1,3,5$
and $1/2^{k}\le m_k\le (5/2)^k$ for $k=2,4$ from their definitions $m_k=\sum_\sigma x_\sigma \sigma^k=\nu\sum_i \sigi^k$.

The map (\ref{mk'=}) reads
\BEQ &&
{m_1}'=\hspace{3mm} \frac{119}{128}+\frac{311 }{320}{m_1}+\frac{5}{16}{m_2}+\frac{1}{8} {m_3}-\frac{1}{8} {m_4}-\frac{1}{20}{m_5}, \nn\\&&
{m_2}'=\hspace{3mm} \frac{119}{128}+\frac{631}{320}{m_1}+\frac{21}{16}{m_2}+\frac{1}{8} {m_3}-\frac{1}{8} {m_4}-\frac{1}{20}{m_5}, \nn\\&&
{m_3}'=\hspace{3mm} \frac{161}{512}+\frac{3489}{1280}{m_1}+\frac{387}{64}{m_2}+\frac{71}{32} {m_3}-\frac{39}{32} {m_4}-\frac{39 }{80}{m_5},
\quad  \nn\\&&
{m_4}'=-\frac{77}{256}+\frac{2227}{640}{m_1}+\frac{377 }{32}{m_2}+\frac{101}{16} {m_3}-\frac{21}{16} {m_4}-\frac{37 }{40}{m_5}, \nn\\&&
{m_5}'=-\frac{12901}{2048}+\frac{10651}{5120} {m_1}+\frac{10865 }{256}{m_2}+\frac{2941 }{128}{m_3}-\frac{1021 }{128}{m_4}-\frac{1341 }{320}{m_5}.
\EEQ
It leaves $C_2$ of eq. (\ref{C2l52}) invariant, and, with $s\to s'=s+1$ mod 6, also $\Is $ of eq. (\ref{hsl52}).
The stable point of the map is the paramagnet described by  $m_1=m_3=m_5=0$, $m_2=\frac{35}{12}$ and $m_4=\frac{707}{48}$.

\section{Summary}
\label{sec7}

Interpretation of quantum mechanics should be based on its touchstone with reality, that is, on the action of an idealized apparatus 
that performs a  large set of measurements on a large set of identically prepared systems. For measurement of the $z$-component of spins-$\frac{1}{2}$
a rich enough model was formulated, the Curie Weiss model for quantum measurement\cite{allahverdyan2003curie}, where the apparatus
consists of an Ising magnet M having itself $N\gg1$ spins-$\frac{1}{2}$, coupled to  thermal harmonic oscillator bath. 
Details of the dynamical solution were summarized and further worked out in ``Opus'' \cite{allahverdyan2013understanding}. 
In order to have an unbiased measurement,
it is required that the Hamiltonian is symmetric under reversal of all spins of M, and that the interaction Hamiltonian
is symmetric under their reversal and reversal of the tested spin.

The purpose of this paper is to construct models to measure the $z$-component of a quantum spin or angular momentum $l\ge 1$,
which takes the values $s=-l,-l+1,\cdots, l$. In order to have an unbiased setup, a $Z_{2l+1}$ invariance is required.
This is achieved by starting from cosines and sines of $2\pi s/(2l+1)$, for the tested spin and the $N$ spins of the magnet, 
which are manifestly invariant under the shift $s\to s+1$ mod $2l+1$.
Shapes for the energy functional and the interaction energy are proposed, which are invariant under 
the shift, and so is and the corresponding entropy.
Since $s$ takes discrete values, the cosines and sines can be expressed in powers $s^k$, $k,=1\cdots,2l$.
For the magnet each of them leads to an order parameter, the first being the magnetization.
The  $Z_{2l+1}$ symmetry now gets coded in a linear map between the order parameters.
The general form of the Hamiltonian, the free energy and the map is worked out for spin $\frac{1}{2}$, $1$, $\frac{3}{2}$, $2$ and $\frac{5}{2}$.
For the spin $1$-case, the thermodynamics is discussed in detail.

To deal with the measurement dynamics, the $x,y,z$ components of each quantum spin of the magnet can be coupled to a harmonic oscillator bath 
like in the spin-$\frac{1}{2}$ case,  which yields the dynamical equations in the early truncation and the subsequent registration periods. 
This subject is presently under study.

In conclusion, the purpose of this work was to support the previous ABN works for interpretation of quantum mechanics based on 
the dynamics of quantum measurement of a spin $\frac{1}{2}$. This goal is achieved by constructing models for spin 1 and larger, that can likewise be investigated dynamically. 
Since it is clear from the ABN works that the measurement dynamics is set by its thermodynamics, it can already be expected  that the new models will exhibit a similar dynamics.
 We have demonstrated that the thermodynamics of the new models is similar in structure to the spin $\frac{1}{2}$ case, be it at the cost of additional order parameters.
The agreement in structure and thermodynamics between the well documented spin $\frac{1}{2}$ model for quantum measurement
and the  present models for larger spin  support the ABN interpretation of quantum mechanics that was put forward previously.


\begin{thebibliography}{-------}
\providecommand{\natexlab}[1]{#1}

\bibitem[Wheeler and Zurek(2014)]{wheeler2014quantum}
Wheeler, J.A.; Zurek, W.H.
\newblock {\em Quantum theory and measurement}; Vol.~15, Princeton University
  Press,  2014.

\bibitem[Allahverdyan \em{et~al.}(2013)Allahverdyan, Balian, and
  Nieuwenhuizen]{allahverdyan2013understanding}
Allahverdyan, A.E.; Balian, R.; Nieuwenhuizen, T.M.
\newblock Understanding quantum measurement from the solution of dynamical
  models.
\newblock {\em Physics Reports} {\bf 2013}, {\em 525},~1--166.

\bibitem[Hepp(1972)]{Hepp1972quantum}
Hepp, K.
\newblock Quantum theory of measurement and macroscopic variables.
\newblock {\em Helvetia Physica Acta} {\bf 1972}, {\em 45},~237--248.

\bibitem[Bell(2001)]{bell2001wave}
Bell, J.S.
\newblock On wave packet reduction in the Coleman--Hepp model. In {\em John S
  Bell On The Foundations Of Quantum Mechanics}; World Scientific,  2001; pp.
  44--49.

\bibitem[Gaveau and Schulman(1990)]{gaveau1990model}
Gaveau, B.; Schulman, L.
\newblock Model apparatus for quantum measurements.
\newblock {\em Journal of statistical physics} {\bf 1990}, {\em
  58},~1209--1230.

\bibitem[Haake and Walls(1987)]{haake1987overdamped}
Haake, F.; Walls, D.F.
\newblock Overdamped and amplifying meters in the quantum theory of
  measurement.
\newblock {\em Physical Review A} {\bf 1987}, {\em 36},~730.

\bibitem[Haake and {\.Z}ukowski(1993)]{haake1993classical}
Haake, F.; {\.Z}ukowski, M.
\newblock Classical motion of meter variables in the quantum theory of
  measurement.
\newblock {\em Physical Review A} {\bf 1993}, {\em 47},~2506.

\bibitem[Allahverdyan \em{et~al.}(2001)Allahverdyan, Balian, and
  Nieuwenhuizen]{allahverdyan2001quantum}
Allahverdyan, A.E.; Balian, R.; Nieuwenhuizen, T.M.
\newblock Quantum measurement as a driven phase transition: An exactly solvable
  model.
\newblock {\em Physical Review A} {\bf 2001}, {\em 64},~032108.

\bibitem[Zeh(1970)]{zeh1970interpretation}
Zeh, H.D.
\newblock On the interpretation of measurement in quantum theory.
\newblock {\em Foundations of Physics} {\bf 1970}, {\em 1},~69--76.

\bibitem[Zurek(2003)]{zurek2003decoherence}
Zurek, W.H.
\newblock Decoherence, einselection, and the quantum origins of the classical.
\newblock {\em Reviews of modern physics} {\bf 2003}, {\em 75},~715.

\bibitem[Schlosshauer(2005)]{schlosshauer2005decoherence}
Schlosshauer, M.
\newblock Decoherence, the measurement problem, and interpretations of quantum
  mechanics.
\newblock {\em Reviews of Modern physics} {\bf 2005}, {\em 76},~1267.

\bibitem[Von~Neumann(2013)]{von2013mathematische}
Von~Neumann, J.
\newblock {\em Mathematische grundlagen der quantenmechanik}; Vol.~38,
  Springer-Verlag,  2013.

\bibitem[Von~Neumann(2018)]{von2018mathematical}
Von~Neumann, J.
\newblock {\em Mathematical foundations of quantum mechanics: New edition};
  Princeton university press,  2018.

\bibitem[Everett~III(1957)]{everett1957relative}
Everett~III, H.
\newblock " Relative state" formulation of quantum mechanics.
\newblock {\em Reviews of modern physics} {\bf 1957}, {\em 29},~454.

\bibitem[DeWitt(1970)]{dewitt1970quantum}
DeWitt, B.S.
\newblock Quantum mechanics and reality.
\newblock {\em Physics today} {\bf 1970}, {\em 23},~30--35.

\bibitem[DeWitt(2015)]{dewitt2015many}
DeWitt, B.S.
\newblock The many-universes interpretation of quantum mechanics. In {\em The
  many-worlds interpretation of quantum mechanics}; Princeton University Press,
   2015; pp. 167--218.

\bibitem[Allahverdyan \em{et~al.}(2003{\natexlab{a}})Allahverdyan, Balian, and
  Nieuwenhuizen]{allahverdyan2003curie}
Allahverdyan, A.E.; Balian, R.; Nieuwenhuizen, T.M.
\newblock Curie-Weiss model of the quantum measurement process.
\newblock {\em EPL (Europhysics Letters)} {\bf 2003}, {\em 61},~452.

\bibitem[Allahverdyan \em{et~al.}(2003{\natexlab{b}})Allahverdyan, Balian, and
  Nieuwenhuizen]{allahverdyan2003quantuma}
Allahverdyan, A.; Balian, R.; Nieuwenhuizen, T.M.
\newblock The quantum measurement process: an exactly solvable model.
\newblock {\em Atti della Fondazione Giorgio Ronchi} {\bf 2003}, p. 719.

\bibitem[Allahverdyan \em{et~al.}(2005{\natexlab{a}})Allahverdyan, Balian, and
  Nieuwenhuizen]{allahverdyan2005quantum}
Allahverdyan, A.E.; Balian, R.; Nieuwenhuizen, T.M.
\newblock The quantum measurement process in an exactly solvable model.
\newblock  AIP Conference Proceedings. American Institute of Physics,  2005,
  Vol. 750, pp. 26--34.

\bibitem[Allahverdyan \em{et~al.}(2005{\natexlab{b}})Allahverdyan, Balian, and
  Nieuwenhuizen]{allahverdyan2005dynamics}
Allahverdyan, A.E.; Balian, R.; Nieuwenhuizen, T.M.
\newblock Dynamics of a quantum measurement.
\newblock {\em Physica E: Low-dimensional Systems and Nanostructures} {\bf
  2005}, {\em 29},~261--271.

\bibitem[Allahverdyan \em{et~al.}(2007)Allahverdyan, Balian, and
  Nieuwenhuizen]{allahverdyan2007quantum}
Allahverdyan, A.E.; Balian, R.; Nieuwenhuizen, T.M.
\newblock The Quantum Measurement Process: Lessons from an exactly solvable
  model. In {\em Beyond the Quantum}; World Scientific,  2007; pp. 53--65.

\bibitem[Allahverdyan \em{et~al.}(2006)Allahverdyan, Balian, and
  Nieuwenhuizen]{allahverdyan2006phase}
Allahverdyan, A.E.; Balian, R.; Nieuwenhuizen, T.M.
\newblock Phase transitions and quantum measurements.
\newblock  AIP Conference Proceedings. American Institute of Physics,  2006,
  Vol. 810, pp. 47--58.

\bibitem[Nieuwenhuizen \em{et~al.}(2014)Nieuwenhuizen, Perarnau-Llobet, and
  Balian]{nieuwenhuizen2014lectures}
Nieuwenhuizen, T.M.; Perarnau-Llobet, M.; Balian, R.
\newblock Lectures on dynamical models for quantum measurements.
\newblock {\em International Journal of Modern Physics B} {\bf 2014}, {\em
  28},~1430014.

\bibitem[Allahverdyan \em{et~al.}(2017)Allahverdyan, Balian, and
  Nieuwenhuizen]{allahverdyan2017sub}
Allahverdyan, A.E.; Balian, R.; Nieuwenhuizen, T.M.
\newblock A sub-ensemble theory of ideal quantum measurement processes.
\newblock {\em Annals of Physics} {\bf 2017}, {\em 376},~324--352.

\bibitem[Allahverdyan \em{et~al.}(2022)Allahverdyan, Balian, and
  Nieuwenhuizen]{allahverdyan2022teaching}
Allahverdyan, A.E.; Balian, R.; Nieuwenhuizen, T.M.
\newblock Teaching quantum measurement: from dynamics to interpretation {\bf
  2022}.

\bibitem[Donker \em{et~al.}(2018)Donker, De~Raedt, and
  Katsnelson]{donker2018quantum}
Donker, H.; De~Raedt, H.; Katsnelson, M.
\newblock Quantum dynamics of a small symmetry breaking measurement device.
\newblock {\em Annals of Physics} {\bf 2018}, {\em 396},~137--146.

\bibitem[Jammer(1966)]{jammer1966conceptual}
Jammer, M.
\newblock {\em The conceptual development of quantum mechanics}; McGraw-Hill,
  1966.

\bibitem[Auff{\`e}ves and Grangier(2016)]{auffeves2016contexts}
Auff{\`e}ves, A.; Grangier, P.
\newblock Contexts, systems and modalities: a new ontology for quantum
  mechanics.
\newblock {\em Foundations of Physics} {\bf 2016}, {\em 46},~121--137.

\bibitem[Auff{\`e}ves and Grangier(2020)]{auffeves2020deriving}
Auff{\`e}ves, A.; Grangier, P.
\newblock Deriving Born’s rule from an Inference to the Best Explanation.
\newblock {\em Foundations of Physics} {\bf 2020}, {\em 50},~1781--1793.

\end{thebibliography}
 \end{document}